\documentclass[12pt,preprint]{aastex}
\usepackage{emulateapj5}
\usepackage{apjfonts}

\begin{document}

\submitted{Accepted by ApJL, 2001 November 13}
\title{Magnetohydrodynamics of Gamma-Ray Burst Outflows}

\author{Nektarios Vlahakis\altaffilmark{1} and Arieh K\"{o}nigl}
\affil{Department of Astronomy \& Astrophysics and Enrico Fermi
Institute, University of Chicago,\\ 5640 S. Ellis Ave., Chicago, IL 60637}
\email{vlahakis@jets.uchicago.edu, arieh@jets.uchicago.edu}

\altaffiltext{1}{McCormick Fellow.}

\begin{abstract}
Using relativistic, axisymmetric, ideal MHD,
we examine the outflow from a disk around a compact object,
taking into account the baryonic matter, the electron-positron/photon
fluid, and the large-scale electromagnetic field.
Focussing on the parameter regime appropriate to $\gamma$-ray
burst outflows, we demonstrate, through exact self-similar solutions,
that the thermal force (which dominates the initial
acceleration) and the Lorentz force (which dominates
further out and contributes most of the acceleration) can
convert up to $\sim 50 \%$ of the initial total energy into
asymptotic baryon kinetic energy. We examine how
baryon loading and magnetic collimation
affect the structure of the flow, including the regime
where emission due to internal shocks could take place.
\end{abstract}


\section{Introduction}
\label{introduction}

Cosmological gamma-ray bursts (GRBs) evidently involve the emission
of the isotropic equivalent of $\sim 10^{51}-10^{54}\ {\rm ergs}$
on a typical time scale of a few seconds in the vicinity of
a newly formed stellar-mass
black hole or rapidly rotating neutron star (see reviews by
\citealt{P99} and \citealt{MRW99}). To avoid strong attenuation of the
$\gamma$-rays by pair-production interactions with lower-energy
photons (the ``compactness'' problem), the Lorentz factor
$\gamma$ of the emitting gas must exceed $\sim 10^2-10^3$. The most
natural way of powering the burst seems to be the
conversion of the kinetic energy of the relativistic outflow
into nonthermal (most likely synchrotron) radiation.
Because of the high initial energy density and
characteristic photon energy, the dynamical evolution of GRBs
has traditionally been modeled in terms of the expansion
of an initially opaque electron-positron ($e^\pm$)
fireball. Although a pure-$e^\pm$ fireball
could in principle attain $\gamma \gtrsim 10^3$ in the optically
thin region where the observed emission must originate, the
fraction of the energy carried by pairs would be only $\sim
10^{-5}$ \citep{GW98}. If, however, the fireball contains
enough baryons (of total mass $M_b$) that it becomes ``matter
dominated'' before it gets optically thin, then most of its
initial (suffix $i$) energy ${\cal E}_i$ is converted into bulk kinetic energy
of baryons
(with asymptotic Lorentz factors $\gamma_\infty \approx {\cal E}_i/M_bc^2$)
instead of escaping as radiation. Baryon loading of
the fireball at the source is expected in view of the highly
super-Eddington luminosity involved, and viable models in fact
face the issue of why it is not so efficient as to
render $\gamma_\infty$ too low (the ``baryon contamination'' problem).

According to currently accepted GRB formation scenarios, the
burst is powered by the extraction of rotational energy from the
central black hole or neutron star, or, alternatively, from the
debris disk left behind when the mass near the origin collapses into a black
hole. In either case, strong ($\gtrsim 10^{14}\ {\rm G}$)
magnetic fields provide the most plausible means of extracting the
energy ${\cal E}_i$ on the burst time scale. If the field were
initially weaker, it would likely be rapidly amplified by differential
rotation or dynamo action to the requisite strength. The magnetic
energy might be dissipated near the origin in a series of
flares, giving rise to a ``magnetic'' fireball (e.g., \citealt{NPP92,MLR93}).
Alternatively, the magnetic field may
have a large-scale, ordered component that could help guide and
collimate the outflow, and, if it is strong enough, also
contribute to its acceleration (e.g., \citealt{U94,T94,MR97,K97}).
Even if the flow is not magnetically driven, the
field may be strong enough to account for the observed
synchrotron emission \citep{SDD01}. Magnetic collimation is, in
fact, consistent with the observational indications for GRB jets
(e.g., \citealt{SPH99}) and could be very helpful in 
reducing the source energy
requirements to plausible values. Furthermore, if the outflow
is largely Poynting flux-dominated, then the implied lower radiative
luminosity near the origin could alleviate the baryon
contamination problem.
Magnetic fields have thus come to be regarded as the favored
means of driving GRB outflows.

The purpose of this Letter is to provide an overview of the
magnetohydrodynamics (MHD) of GRB outflows, clarifying the
relationship between the thermal (fireball) and Lorentz accelerations,
identifying the parameter
regimes where qualitatively different behaviors are expected,
and demonstrating that the observationally inferred properties of
these outflows can be attributed to magnetic driving. A
detailed exposition of the semianalytic solutions on which this overview 
is based is given in N. Vlahakis \& A. K\"onigl, in preparation (hereafter VK).

\section{The MHD Description}
\label{MHD}

For the purpose of illustration, we assume that the outflow
originates in a debris disk around a stellar-mass black hole
(e.g., \citealt{MR97,FWH99}). In light of the considerations of
\S 1, we concentrate on
the case of a magnetically driven, axial jet that has a sufficiently high
baryon loading to insure that it is matter dominated 
when it becomes optically thin (see \S \ref{baryon}), and in
which a significant fraction of the Poynting flux is eventually
converted into kinetic energy. As discussed by Spruit et
al. (2001), such configurations can be
validly described by the MHD approximation on the spatial scales of interest,
and, furthermore, they are not expected to dissipate a
substantial amount of magnetic energy along the
way. Correspondingly, we adopt the equations of relativistic,
ideal MHD to describe the flow. We anticipate that, near the
origin, the thermal energy associated with the radiation and
$e^\pm$ pairs is nonnegligible and that the
optical depth is large enough to ensure local thermodynamic
equilibrium. We therefore assume that the gas (consisting of
baryons with their neutralizing electrons as well as of photons
and pairs) evolves adiabatically with a polytropic index of 4/3.
Assuming a quasi-steady poloidal magnetic flux function $A$ and
changing variables from $(A\,,\ell\,,t)$ to $(A\,,\ell\,,s=ct-\ell)$,
with $\ell$ the arclength along a poloidal fieldline,
it can be shown that all terms with derivatives with respect
to $s$ are negligible when the flow is highly relativistic.
As is elaborated on in VK,
the equations are then effectively time independent
and the motion can be described as a frozen pulse whose internal
profile is specified through the variable $s$ (see also Piran 1999).
We discuss time-dependent effects in \S \ref{application}.

Assuming also axisymmetry, the full set of MHD equations can
be partially integrated to yield several field-line constants: 
the mass-to-magnetic flux ratio $\Psi_A(A\,,s)$,
the field angular velocity $\Omega(A\,,s)$
(which equals the matter angular velocity
at the footpoint of the field line in the disk),
the total (kinetic + magnetic) specific angular momentum $L(A\,,s)$,
the total energy-to-mass flux ratio $\mu(A\,,s)c^2$,
and the adiabat $Q(A\,,s)\equiv P/ \rho_0^{4/3}$ (where $\rho_0$ and $P$
are, respectively, the rest-mass density and total
gas pressure measured in the fluid frame). Two integrals remain to be
performed, involving the Bernoulli and transfield force-balance
equations. There are correspondingly two unknown functions,
which we choose for convenience to be $x\equiv \varpi \Omega / c$, the radial
distance of the field line [in cylindrical coordinates $(z\,,\varpi\,,\phi)$]
in units of the ``light cylinder'' radius, and the ``Alfv\'enic'' Mach number
$M\equiv \sqrt{4 \pi \rho_0 \xi} ( \gamma V_p / B_p)
=\Psi_A \sqrt{\xi / 4 \pi \rho_0}$ (where $\xi$ is the
enthalpy-to-rest energy ratio, and where $V_p$ and $B_p$ are
the poloidal components of the velocity and magnetic field,
respectively, measured in the central object's frame).
The solutions derived in VK are obtained under the most general
ansatz for radial self-similarity [in spherical coordinates
$(r\,,\theta\,,\phi)$], in which the shape $r(\theta,A)$ of the
poloidal field lines is given as a product of a function of $A$
times a function of $\theta$:
$r={\cal F}_1(A) {\cal F}_2 (\theta)$
(or, equivalently, $A={\cal A}[\varpi/G(\theta)]$; \citealp{VT98}).

Initially, the outflow evolves as a fireball: it expands
isotropically in the comoving frame, with the large-scale
field only acting as a guide (e.g., M\'{e}sz\'{a}ros et al. 1993).
If the baryon loading is large enough, this continues
until all the thermal energy of the radiation field and the
pairs is transformed into kinetic energy of the baryons.
Subsequently, the Lorentz force converts Poynting flux into
baryon kinetic energy, and it also collimates the flow.
The Lorentz force remains dominant (implying that the field is
force free) all the way to the point where the kinetic
energy of the baryons becomes comparable to the electromagnetic energy.
We assume that the asymptotic shape of the flow is cylindrical and that
the final stage is characterized by an equipartition between the
Poynting and kinetic energy fluxes. (The latter assumption is
motivated by the exact ``cold'' solutions of \citealt{LCB92}.)
The transition from a parabolic to a cylindrical shape occurs far downstream 
from the Alfv\'{e}n point ($\theta \ll 1$), where the ratio of
the kinetic energy flux $\xi \gamma^2 \rho_0 c^3$ to the
Poynting flux $c E B_{\phi} / 4 \pi$ (where $E$ is the electric
field amplitude) equals $M^2/x^2$. 
If $\vartheta$ is the (small) opening half-angle of the outflow, then
$\mid {\bf \nabla} A \mid \sin(\theta-\vartheta)=
{\partial A}/{\partial r}$.
Substituting this into the Bernoulli equation, we find
the slope of the field lines
\begin{equation}\label{slope}
\left(\frac{d\ln \varpi}{d \ln z}\right)_A
=\frac{\vartheta}{\theta}
=1-\left(1+\frac{M^2}{x^2}\right)
\frac{\sigma}{\mu}
\frac{r}{A} 
\left(\frac{\partial A}
{\partial r}\right)_{\theta}\,,
\end{equation}
where $\sigma\equiv A \Omega^2 /( c^3 \Psi_A)$ is the magnetization parameter.
Equipartition between the Poynting and kinetic energy fluxes $(M^2=x^2)$
in the final cylindrical stage $(d\varpi=0)$ means that
$r\partial A(r\,,\theta)/ A \partial r=\mu/ (2 \sigma)$.
Substituting this into equation (\ref{slope}) and specializing to the
force-free $(M^2 \ll x^2)$ parabolic regime, gives
$d \ln z / d\ln \varpi = 2$, so the shape of the
force-free field lines is
$z/\varpi_i =( \varpi / \varpi_i )^2 / (2 \tan \vartheta_i)$,
with $\vartheta_i$ the initial opening half-angle
(cf. \citealt{C95}).

\section{Scaling Laws}
Beyond the light surface (which is close to the Alfv\'{e}n
surface, since the flow is nearly force free), the large-scale
electromagnetic field is given by
\begin{equation}
\label{fields}
(B_z,B_{\varpi},B_{\phi},E_z,E_{\varpi},E_{\phi})=
\frac{\mu c \Psi_A}{x}
\left(\frac{1}{x},\frac{\varpi}{2 x z},-1,
\frac{\varpi}{2 z},-1, 0\right).
\end{equation}
The comoving field amplitude satisfies $B_{\rm co}^2=B^2-E^2$
(a Lorentz invariant) and $B_{\rm co}^2 \approx B_p^2+B_{\phi}^2 /
\gamma^2 \ll B_{\phi}^2$, which together imply $B_{\phi} \approx E_{\varpi}$
(or, equivalently, $L \Omega \approx \mu c^2$).

We approximate the outflow as a pair of shells that move in opposite
directions from the debris disk, each having an initial
meridional cross section $(\Delta z)_i \times (\Delta \varpi)_i $, with
$(\Delta \varpi)_i=10^6 (\Delta \varpi)_{i,6} \ {\rm cm}$
comparable to $\bar{\varpi}_i=10^6 \bar{\varpi}_{i,6} \ {\rm cm}$
(the mean radius of the debris disk)
and $(\Delta z)_i \approx c \, \Delta t = 3
\times 10^{11}\Delta t_{1}$ cm (where $\Delta  t = 10 \,\Delta
t_{1}\ {\rm s}$ is the total duration of the burst).
The total baryonic mass of outflowing matter is
$M_b=4 \pi \bar{\varpi}_i (\Delta \varpi)_i (\Delta z)_i \gamma_i \rho_{0i}=
2 \times 10^{-7} \gamma_i \rho_{0i,2} \bar{\varpi}_{i,6} (\Delta\varpi)_{i,6}
\Delta t_{1} \, M_{\sun}\,,$
where $\rho_{0i} = 10^2 \rho_{0i,2} \ {\rm g\ cm^{-3}}$.
The total energy is ${\cal E}_i = \mu M_b c^2=
4 \times 10^{51} (\mu/10^4) \gamma_i \rho_{0i,2}
\bar{\varpi}_{i,6} (\Delta\varpi)_{i,6} \Delta t_{1}\ {\rm
ergs}$, and initially it resides
predominantly in the electromagnetic field; the initial thermal
energy (associated with the enthalpy
of the photons and pairs) is $\xi_i M_b c^2 = (\xi_i / \mu) {\cal E}_i$.

The constancy of the mass-to-magnetic flux ratio
$\Psi_A=4 \pi \rho_0 \gamma V_p / B_p$
and equation (\ref{fields}) imply
\begin{equation}\label{continuity}
\gamma \rho_0 \varpi^2 = \gamma_i \rho_{0i} \varpi_i^2\, .
\end{equation}
If $\Theta$ is the temperature in units of the electron rest
energy, then, by the conservation of specific entropy along the
flow, $P/ (\rho_0 \Theta) = const$, or (since $P \propto \Theta^4$)
\begin{equation}\label{entropy}
\rho_0= \rho_{0i} (\Theta/\Theta_i)^3 \, .
\end{equation}
Energy flux conservation, in turn, gives
\begin{equation}\label{energy}
\xi \gamma - (\varpi \Omega B_{\phi}/\Psi_A c^2) = \mu\, .
\end{equation}
So long as $\xi>1$ (the fireball phase), $\xi \propto
\Theta$. Also, by equation (\ref{energy}), $\xi \gamma$ is
constant since the specific
Poynting flux does not vary along the flow for a force-free field.
Equations $(\ref{continuity})-(\ref{energy})$ thus imply
\begin{equation}\label{scale}
\gamma = \gamma_i (\varpi/\varpi_i)
\,,\,\,\,\,\,
\rho_0 = \rho_{0i} (\varpi_i/\varpi)^3
\,,\,\,\,\,\,
\Theta = \Theta_i (\varpi_i/\varpi)\,.
\end{equation}
These scalings are the same as in a spherically symmetric,
hydrodynamic (HD) fireball, except that the cylindrical radius $\varpi$
replaces the spherical radius $r$ in the magnetically guided case.

Beyond the point where $\xi=1$, magnetic acceleration takes over
and the Lorentz factor continues to increase until the
cylindrical regime ($\varpi=\varpi_{\infty}$) is reached.
The magnetic acceleration is associated with a small
(linear in $\varpi$) deviation of the Poynting flux from its force-free value,
\begin{equation}\label{poynting}
-(\varpi \Omega B_{\phi}/\Psi_A c^2)=
\mu - (\varpi/{\varpi_i})\, .
\end{equation}
By combining equation (\ref{poynting}) with equations
(\ref{energy}) and (\ref{continuity}), it is seen that the
scalings of $\gamma$ and $\rho_0$ with $\varpi$ given by
equation (\ref{scale}) are maintained also in the magnetic
acceleration zone. In fact, since the density in the central object's frame is
$\gamma \rho_0 = \gamma_i \rho_{0i} (\varpi_i
/\varpi)^2$ and the $\varpi$-width of the shell (given the
parabolic shape of the field lines) evolves as
$\Delta \varpi / (\Delta \varpi)_i = \varpi/\varpi_i$, mass
conservation implies that, while the flow continues to expand
isotropically in the comoving frame,
the width of the shell remains constant
[$\Delta z=(\Delta z)_i$] in the central source's frame
(consistent with the frozen-pulse approximation).

\section{Baryon Loading}
\label{baryon}

Two parameters determine the basic character of the flow:

\noindent 
The first is the 
initial enthalpy of the flow in units of the
rest energy of the baryonic component,
$\xi_i=400 \, \Theta_i^4 (\rho_{0i,2})^{-1}$,
where we approximate the initial radiation field by a blackbody
distribution and the $e^\pm$ pairs by a Maxwellian distribution.

\noindent
The second is the 
initial Poynting flux in units of the baryon rest-
energy flux. As we focus on Poynting flux-dominated outflows,
this ratio is $\mu- \xi_i \approx \mu$.

Initially the flow is optically thick.
The distance where it becomes optically thin
is determined by $\xi_i$. The pertinent regimes of this
parameter were previously analyzed for the case of a radially
expanding HD fireball \citep[see][]{P99}.
We now adapt this analysis to the present (nonradial) geometry.

The opacity has two contributions: one due to pairs ($\tau_{\pm}$)
and another due to the electrons that neutralize the baryons ($\tau_b$).
As the flow expands, its temperature decreases.
When $\Theta \approx 0.04$, the pair number density becomes
negligible and the opacity from there on is determined
by the neutralizing electrons. For the radiation-dominated case,
$\tau_b<1$ at that point, corresponding to $\xi_i > \xi_{i\pm}$ (with
$\xi_{i\pm}=2.5 \times 10^7 \Theta_i^2 (\Delta \varpi)_{i,6}$).
If $\xi_i < \xi_{i\pm}$, the flow continues to
be optically thick until $\tau_b=1$. If at that point the
radiation energy density exceeds the baryon rest-energy
density, then one is still in the radiation-dominated
regime. This corresponds to $\xi_i > \xi_{ib}$, with
$\xi_{ib}=6 \times 10^3 \Theta_i^{4/3} (\Delta \varpi)_{i,6}^{1/3}$.
For $\xi_i < \xi_{ib}$, the flow is matter dominated when it
becomes optically thin. Finally, $\xi_i\approx 1$ correspond to the cold limit.
Since magnetic acceleration now plays a role, the outflow
is not necessarily nonrelativistic in this
limit: only if $\mu \approx 1$ also holds will the flow be Newtonian.
The most plausible regime for GRB outflows is $1<\xi_i < \xi_{ib}$,
and we adopt it from here on.

\section{Structure of the Flow}\label{structure}
As the flow moves outward
it passes through the following
points (see Fig. \ref{fig1}):
\\
1.
$\varpi_1 =\varpi_i $\,: origin of the outflow,
$\gamma_i \approx 1$.
\\
2.
$\varpi_2 =\varpi_i \sqrt{3/2}$\,:
slow-magnetosonic surface, $V_p\approx c/\sqrt{3}$.
\\
3.
$\varpi_3 =c/ \Omega$\,:
Alfv\'{e}n ($\approx$ light) surface.
\\
4.
$\varpi_4 = (c/ \Omega) \sqrt{\mu / \xi_i}$\,:
classical fast-magnetosonic surface,
$\gamma V_p \approx \sqrt{(B^2-E^2)/(4 \pi \rho_0 \xi)}$.
Unlike the purely radial case
\citep{M69}, this point is located at a finite distance from the origin.
The bulk of the magnetic
acceleration occurs downstream from this point through the
``magnetic nozzle'' mechanism (Li et al. 1992). In essence, this term
is a shorthand for the fact that an MHD flow can continue to
accelerate until it crosses the modified fast-magnetosonic singular
surface.\footnote{The modified (and {\em not} the classical)
fast-magnetosonic surface
is a singular surface for the steady MHD equations
when one solves simultaneously the Bernoulli and transfield
equations (e.g., \citealp{B97}).}
This effect is not purely relativistic: it is manifested in an
exact solution of the nonrelativistic MHD equations where all the singular
surfaces (including the modified-fast one) are crossed
\citep{V00}, and there, too, most of the acceleration occurs
downstream of the classical fast-magnetosonic surface (but {\em
upstream} of the modified-fast surface).
\\
5.
$\varpi_5 = 25\, \Theta_i \varpi_i$\,: $\tau_{\pm}=1$.
\\
6.
$\varpi_6 =\xi_i \varpi_i$\,:
end of thermal acceleration, $\gamma \approx \xi_i$.
The entire initial thermal energy of the photons and pairs has
by this point been converted into baryon kinetic energy.
Magnetic acceleration effectively starts here.
\\
7.
$\varpi_7 =
5 \times 10^3 (\Delta \varpi)_{i,6}^{1/2}
\rho_{0i,2}^{1/2} \varpi_i$\,:
$\tau=\tau_b=1$, the flow becomes
optically thin to Compton scattering.
\\
8.
$\varpi_8 = \varpi_{\infty} \approx (\mu / 2) \varpi_i$\,:
cylindrical flow regime. Near-equipartition between the
Poynting and the baryon kinetic-energy
fluxes is attained, with $\gamma = \gamma_\infty \approx \mu/2$.

\begin{center}
\plotone{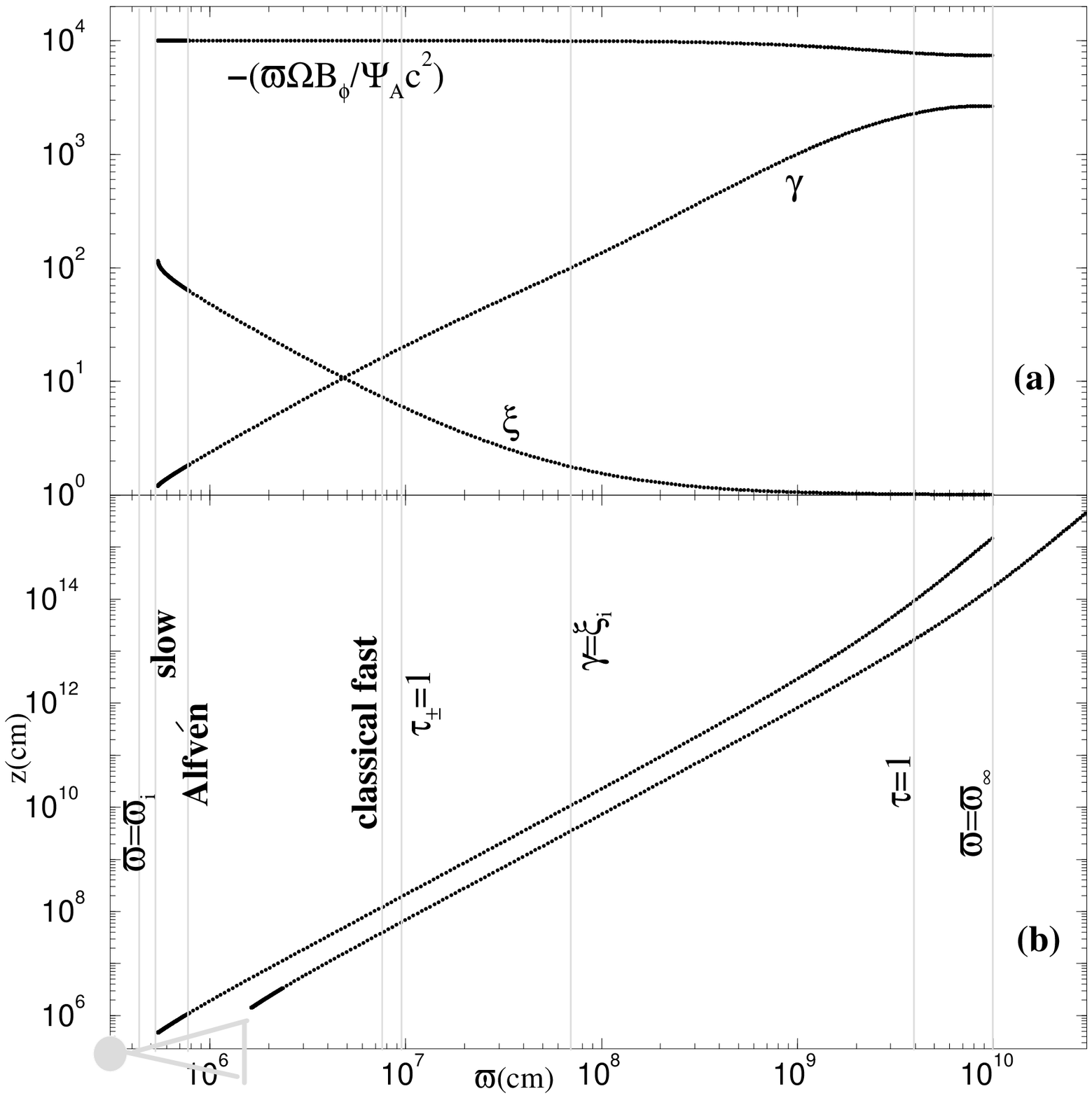}
\figcaption[]
{Exact self-similar solution of
the relativistic MHD equations, illustrating the GRB outflow
structure discussed in this Letter. ($a$) The Lorentz factor
$\gamma$, the ratio $\xi$ of the enthalpy to the rest energy, and the
ratio of the Poynting flux to the rest-energy flux ({\it top}\/ curve)
are shown as functions of $\varpi$, the distance from the axis of rotation,
along the innermost field line. ($b$) The meridional projections
of the innermost and outermost field lines are shown on a
logarithmic scale, along with a sketch of the
black-hole/debris-disk system.
The vertical lines mark the positions of the
various transition points listed in \S \ref{structure} along the
innermost field line.
\label{fig1}}
\end{center}

The inequalities $\varpi_1 < \varpi_2 < \varpi_3 < \varpi_4 <\varpi_8$
always hold, as do, in the regime of interest, $\varpi_5 <
\varpi_7$ (valid for $\xi_i < \xi_{i \pm}$) and
$\varpi_6 < \varpi_7$ (valid for $\xi_i < \xi_{i b}$). The
validity condition for the inequality $\varpi_7 \lesssim \varpi_8$, which
expresses the requirement that the flow be optically thin
to Compton scattering in its final (cylindrical) stage,
corresponds to an upper bound on the baryon loading:
$\rho_{0i,2} \lesssim
\left(\mu / 10^4\right)^2  (\Delta \varpi)_{i,6}^{-1}$
[or $\rho_{0i,2} \lesssim 0.4 (\Delta \varpi)_{i,6}^{-1}
\bar{\varpi}_{i,6}^{-2/3} (\Delta t_1)^{-2/3}
({\cal E}_i / 10^{51} {\rm ergs})^{2/3}$]. We note, however,
that the increase of $\gamma$ with $\varpi$
is slower than linear as the cylindrical regime is
approached (see Fig. \ref{fig1}), which implies that $\varpi_{\infty}$
is larger than $(\mu/2) \varpi_i$ and
the actual upper bound on the initial
baryon mass density is larger than the
analytic estimate (typically by a factor of a few).

All of the above points are indicated in Figure \ref{fig1}, where an
illustrative solution of the steady, relativistic MHD
equations is presented. The figure demonstrates the validity of
the scaling $\gamma \approx \varpi/ \varpi_i$ over several
decades in $\varpi$ and the separation
of the thermal ($\gamma<\xi_i$) and magnetic ($\gamma>\xi_i$) acceleration
regimes. It also manifests the significant collimation from the
comparatively large initial
opening half-angle ($\vartheta_i \approx 24 \degr$)
to a very nearly cylindrical geometry (attained on scales
$\gtrsim 10^{14}\ {\rm cm}$ from the origin).
Equation (\ref{poynting}) for the Poynting flux is also
verified. It is, furthermore, seen that approximate
equipartition ($\gamma \approx
-\varpi \Omega B_{\phi}/\Psi_A c^2$) holds during the final
phase of the flow.

\section{Application to GRB{\scriptsize S}}
\label{application}

To demonstrate how the model outflows discussed in \S
\ref{structure} could account for observed GRBs, we adopt as
fiducial source parameters a $\gamma$-ray fluence of $\sim
10^{-5}\ {\rm ergs\ s^{-1}}$ and a distance $\sim 3\ {\rm Gpc}$,
which imply an isotropic equivalent energy of $\sim 10^{53}\ {\rm ergs}$.
If the $\gamma$-ray emitting material
is in fact confined to a pair of cones of opening half-angle $\sim 3\degr$, 
then the actual radiated energy is $\sim 1.5 \times 10^{50}\ {\rm ergs}$.
As we noted in \S 1, the emission is believed to be powered by
the kinetic energy of the relativistic outflow, and the most
common interpretation is in terms of internal shocks produced by
the collision of overtaking shells \citep[see][]{P99}. Assuming
that the kinetic-to-radiative energy conversion efficiency is $\sim 10\%$
and that the magnetic stresses can transfer as much as $\sim
1/2$ of the initial energy into baryon kinetic energy, we infer
${\cal E}_i \approx 3 \times 10^{51}\ {\rm ergs}$. 
If the energy is deposited in
the form of a Poynting flux over $\Delta t \approx 10\ {\rm s}$,
then the field at the origin must be
$\sim ({\cal E}_ic/\bar{\varpi}_i^3 (\Delta \varpi)_i \Delta t \Omega^2)^{1/2}
\approx 3 \times 10^{14}\ {\rm G}$
for $\bar{\varpi}_i \approx (\Delta \varpi)_i \approx
10^6\ {\rm cm}$ and $\Omega \approx 10^4\ {\rm s^{-1}}$
(representative of a debris disk that extends
to the last stable orbit of a rotating solar-mass black hole). 
To satisfy the constraint $\varpi_7 \lesssim \varpi_8$ (see \S
5), the maximum baryonic rest-mass density is $\rho_{0i,2} \approx 1$.
In the solution presented in Figure 1, we adopted as a plausible
temperature $\Theta_i \approx 1/\sqrt{2}$,
corresponding to $\xi_i \approx 100$.

The variability properties of the observed GRBs
have been interpreted in terms of $N=100\ N_{2}$ distinct shells
that are ejected with slightly different
Lorentz factors and whose subsequent collisions give rise to the
``internal'' shocks responsible for the $\gamma$-ray emission
(e.g., \citealt{P99}). Adopting $\gamma \propto \varpi \propto \sqrt{z}$,
one may integrate the equation of motion for each shell
and show that two neighboring shells starting with $\Delta \gamma_i \sim 1$
will not collide for as long as they move in the flow acceleration zone.
Only after the shells reach the constant-velocity, cylindrical
flow regime, will the two shells collide (at a height $z_f \approx
\gamma_{\infty}^2 (\Delta z)_i /N
= 7.5 \times 10^{16} (\mu/10^4)^2 \Delta t_{1} / N_{2}\ {\rm
cm}$, assuming an initial separation $\sim (\Delta z)_i /N$). A
lower bound on the Lorentz factor in this region can be obtained
from the requirement that the optical depth to photon-photon pair production
be less than 1: $\gamma \gtrsim 10^3 \epsilon_{t,1}^{1/4}
N_2^{1/4} (\Delta t_1)^{-1/4} (1+z_r)^{1/4}
H_{65}^{-1/2}$, where $\epsilon_{t,1}\ {\rm GeV}$ is the energy of a test
photon, $65 H_{65}\ {\rm km\ s^{-1}\ Mpc^{-1}}$ is the Hubble constant,
and $z_r$ is the redshift \citep{WL95}. The solution exhibited in Figure
\ref{fig1} demonstrates that magnetically accelerated flows can
attain the requisite high values of $\gamma$.

Although our analysis can be generalized to arbitrary magnetic
field configurations, we have concentrated on the self-similar form
for which we have exact
solutions of the MHD equations
(see VK). Real GRB outflows
evidently have a finite opening half-angle
(estimated to be at least $5 \degr$ on
average; M\'{e}sz\'{a}ros et al. 1999), and
thus are not accurately represented by our asymptotically
cylindrical solutions. This discrepancy should not, however,
affect our basic conclusions about the robustness of the
magnetic acceleration mechanism for these flows.
It remains a challenge for future work to find exact solutions
with conical asymptotics.
We note in this connection
that, although the outflowing ``shells'' conceivably do not fill the entire
solid angle into which they are ejected, the internal shock
mechanism requires that multiple shells be ejected along any
given direction (e.g., \citealt{KP00}). The guiding property of a
magnetic field in an MHD-driven outflow provides a natural
physical basis for this picture.

In conclusion, we have shown that ordered magnetic fields can
transform up to $\sim 50\%$ of the energy deposited by the
central source of a GRB into kinetic energy of
a collimated flow of baryons with $\gamma \sim 10^3$.
This energy, in turn, may be converted into radiation by internal
shocks.

\acknowledgements
We thank J. Granot for helpful conversations.
This work was supported in part by NASA grant
NAG 5-9063 and by DOE under Grant No. B341495.

\end{document}